# Wireless Computing and IT Ecosystems

William R Simpson

*Institute for Defense Analyses, 4850 Mark Center Drive, Alexandria, Virginia 22311 USA,*
rsimpson@ida.org

**Abstract.** We have evolved an IT system that is ubiquitous and pervasive and integrated into most aspects of our lives. Many of us are working on 4th and 5th level refinements in efficiency and functionality. But, we stand on the shoulders of those who came before and this restricts our freedom of action. The prior work has left us with an ecosystem which is the living embodiment of our state-of-the-art. While we work on integration, refinement, broader application and efficiency, the results must move seamlessly into the ecosystem. Fundamental concepts are being researched in the lab and may rebuild the world we all live in, until that happens, we must work within the ecosystem.

## 1   Introduction

Today, a combination of standards, applications, methodologies and interfaces has evolved into an ecosystem call web 2.0. Although the name would indicate a new version of the World Wide Web, it does not refer to any technical specification, but rather to the way software developers and end-users use the web today [1]. O'Reilly Media publisher Dale Dougherty coined the phrase Web 2.0 In 2004 [2]. Searching the web will not provide a clear definition of web 2.0. The author takes the term as a maturity level of processes, standards, and applications that form an ecosystem around which other applications and processes cannot diverge significantly or there cost to adopt them will be excessive.

## 2   The Layers of an Ecosystem

The ecosystem is a living, breathing embodiment of an IT system of systems that is ubiquitous and pervasive and integrated into most aspects of our lives. It adapts to environment changes and enforces a measure of conformance by its very existence.

### 2.1   Ecosystem Layers

Describing an architecture that conforms to these concepts requires specific elements and standards, a partial sampling is given in references [3] – [7] to ensure the processes are built into systems. The layers are similar to the layers in an onion as shown if figure 1.

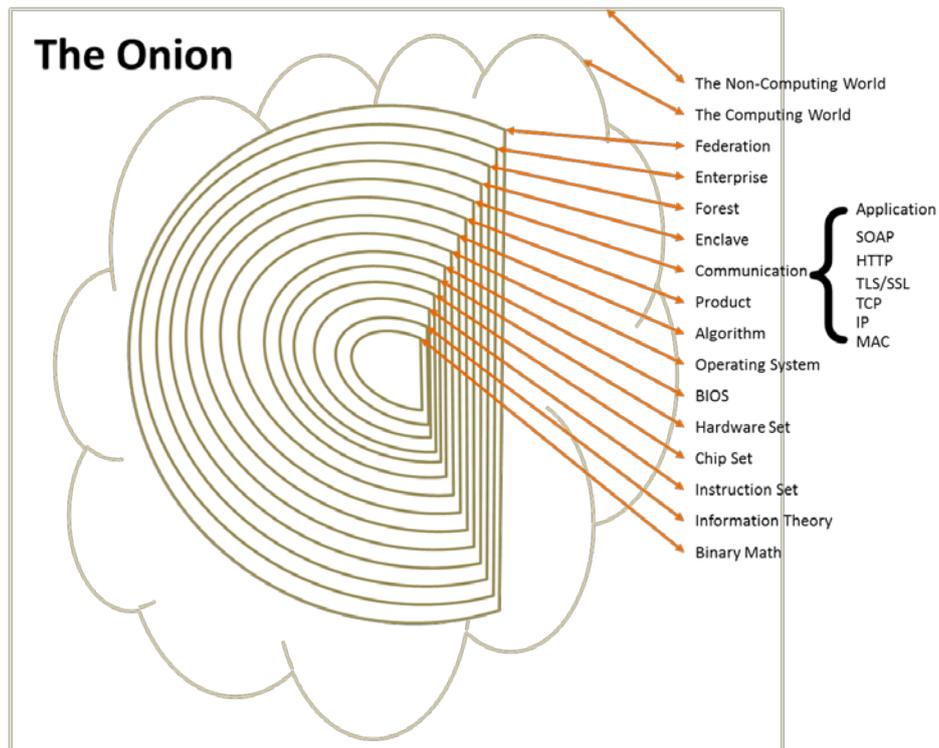

**Fig. 1**. The Layers of the Ecosystem

### 2.2  Layers within layers

The layers of figure 1 are exemplary and have not been blessed by any organization. Certain of the sub-layers have. In figure 1 we have broken out the sub-layers of one category – communications – and apply the open systems interconnect reference model [8] for its constituent sub-layers shown below in figure 2. The model has been used for the definition of end points and protocols for communications processes.

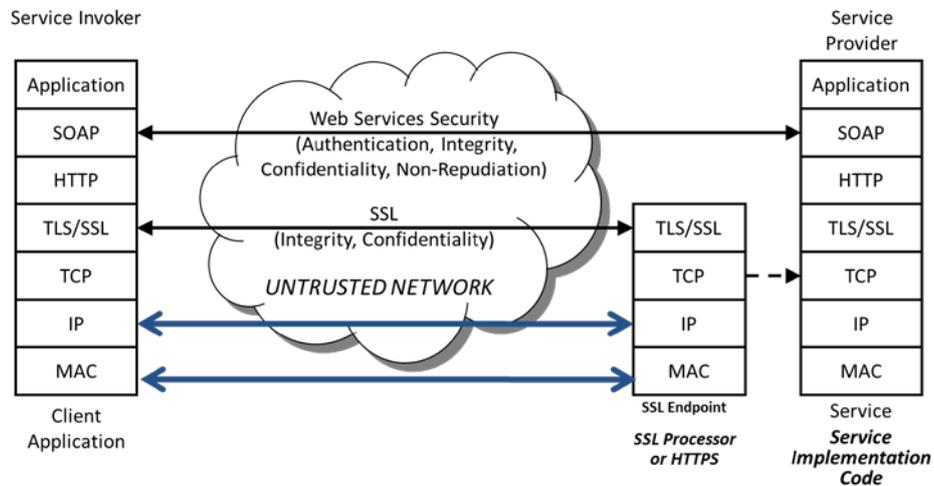

**Fig. 2.** Secure Communication OSI Reference Model

## 3  Working within the Layers

Wherever you work in this ecosystem, you must be concerned with:

- Functionality- A Manet must provide routing of messages to members of its network.
- Efficiency – A pheromone based routing method must be more efficient than a priority-based or round-robin routing to make it worth the added complexity.
- Security – Even in ad-hoc Manet based networks we must protect our own assets and monitor nefarious behaviors. The following are some threat mitigation and elimination that may be considered:
    a. Traditional threats
        i. Threats against the OS and Network interfaces
        ii. Authentication, Authorization attacks
        iii. Man in the middle attacks inside / outside enterprise both on the platform and on all comm. Links
        iv. Configuration setup and Change based attacks,
        v. Mobile software attacks
        vi. Integrity Attacks [change security attributes, metadata record content]
        vii. Design and Implementation threats
            1. Buffer overflow, Static Analysis, Type checking
            2. Web service attacks [cross site scripting, ] others
            3. Threats on Caches, Replays

       b. Data at rest and data in Motion threats
       c. Cryptographic key PKI private key and session key threats [Java]
       d. Attacks on Virtualization [Minimal]
       e. Attack on Inter Forest / Enclave Trust, TPM and Others …
       f. Threats from Black Hat

Integrity – software, data, credentials all need to be validated as to source and tamper resistance.

Standard Interfaces – unless we are in a standalone environment we must be able to communicate with other elements in the ecosystem.

Records Management – Record pertinent events for monitoring and forensics:
       a. Security violation
       b. Performance issues
       c. Fault recoveries
       d. Health Monitoring
       e. Behavioral Monitoring
       f. Others dealing with the functionality
       g. Etc.

If you are not working in all of these areas, your work may not survive the laboratory.

## 4     Vulnerabilities in Software

Vulnerabilities occur at all levels of software development. Most of the software developed for wireless communication are coded to the middle of the OSI communications model. Vulnerabilities are tracked and aggregated according to the software level, and the weaknesses in coding by the Common Weakness Enumeration (CWE) [9] which describes itself as follows:.

> "CWE - International in scope and free for public use, CWE provides a unified, measurable set of software weaknesses that is enabling more effective discussion, description, selection, and use of software security tools and services that can find these weaknesses in source code and operational systems as well as better understanding and management of software weaknesses related to architecture and design."

A summary by type is provided below as derived from the CWE for middleware coding vulnerabilities. The broader category includes a broad range of middleware software including Oracle data base code and Apache HTTP server code, but can be used as a gauge for software vulnerabilities that arise during normal coding, using standard coding reviews. You could expect the numbers to be considerably higher for software developed within academic experiments and experimental code with less stringent development activities.

Table 1 Weakness Enumerations

| | Common Weakness Enumeration for Non-Website Vulnerabilities |
|---|---|
| 1 | Improper Restriction of Operations within the Bounds of a Memory Buffer (CWE-119) – 32% |
| 2 | Improper Control of Generation of Code ('Code Injection') (CWE-94) – 17% |
| 3 | Resource Management Errors (CWE-399) – 10% |
| 4 | Improper Input Validation (CWE-20) – 9% |
| 5 | Numeric Errors (CWE-189) – 8% |
| | **Middleware Coding Vulnerability Types** |
| 1 | Improper Input Validation (CWE-20) – 11% |
| 2 | Resource Management Errors (CWE-399) – 11% |
| 3 | Improper Control of Generation of Code ('Code Injection') (CWE-94) – 9% |
| 4 | Numeric Errors (CWE-189) – 9% |
| 5 | Improper Restriction of Operations within the Bounds of a Memory Buffer (CWE-119) – 7% |

# 8 Summary

This paper has reviewed the eco-system that most research will eventually have to live within. The ultimate goal is to have the algorithms coded into products and deployed into an enterprise or web service environment. Shortly after your research phase and as you begin product development remember these factors. At whatever layer you are working, you are responsible for; Functionality, Efficiency, Security, Integrity, Standard Interfaces, Records Management(for forensics), Etc. Shortfalls will have to be covered in the next level in the ecosystem, often at a penalty in performance and efficiency. These items should be "baked in" not "added on". Eliminate Common vulnerabilities. Your algorithm (no matter how efficient) may not become productized or institutionalized if you miss these factors.